\documentclass[a4paper,9pt]{article}
\usepackage{graphicx,url}
\usepackage[all]{xy}
\usepackage{cite}
\usepackage[top=1.5cm,bottom=1.5cm,left=2.5cm,right=2.5cm]{geometry}
\usepackage[utf8]{inputenc}
\usepackage[colorlinks = true,
            linkcolor = black,
            urlcolor  = blue,
            citecolor = black,
            anchorcolor = black]{hyperref}
\usepackage{authblk} 

\usepackage{multicol} 
\usepackage{etoolbox}
\patchcmd{\thebibliography}{\section*{\refname}}
   {\begin{multicols}{2}[\section*{\refname}]}{}{}
\patchcmd{\endthebibliography}{\endlist}{\endlist\end{multicols}}{}{}

\patchcmd{\thebibliography}{\section*{\refname}}{}{}{}

\usepackage{multicol,caption,setspace} 
\usepackage{etoolbox}

\usepackage[usenames, dvipsnames]{color}
\usepackage[normalem]{ulem}

\usepackage{amsfonts}
\usepackage{amssymb}
\usepackage{amsbsy} 
\usepackage{amsmath}
\usepackage{mathrsfs}
\usepackage{dsfont}
\usepackage{bbm}
\usepackage{subfigure}

\usepackage{float}
\usepackage{flushend}

\usepackage{scrextend}

\newenvironment{Figure}
  {\par\medskip\noindent\minipage{\linewidth}}
  {\endminipage\par\medskip}

\newcommand{\comment}[1]{}
\definecolor{marina}{rgb}{1, 0, 0.5}

\newcommand{\vect}[1]{\boldsymbol{#1}}
\title{Appropriate kernels for Divisive Normalization \\ explained by Wilson-Cowan equations\footnote{Parts of this work have been presented at MODVIS-18, and at the Celebration of Cowan's 50th Anniv. Univ. Chicago}}
\date{\vspace{-5ex}}
\author[1]{\vspace{-0.3cm}J. Malo\footnote{Correspondence: jesus.malo@uv.es}}
\author[2]{M. Bertalmio\vspace{-0.3cm}}
\affil[1]{\vspace{-0.00cm}\small{Image Processing Lab. Parc Científic, Universitat de València, Spain}}
\affil[2]{\vspace{-0.00cm}Dept. Tecnol. Inf. Comunic., Universitat Pompeu Fabra, Barcelona, Spain}

\begin{document}

\maketitle
\thispagestyle{empty}
\renewcommand{\baselinestretch}{1.2}
\vspace{-0.0cm}

\begin{abstract}

The interaction between wavelet-like sensors in Divisive Normalization is classically described through Gaussian kernels
that decay with spatial distance, angular distance and frequency distance.
However, simultaneous explanation of
(a)~distortion perception in natural image databases and
(b)~contrast perception of artificial stimuli
requires very specific modifications in classical Divisive Normalization.
First, the wavelet response has to be high-pass filtered before the Gaussian interaction
is applied. Then, distinct weights per subband are also required after the Gaussian interaction.
In summary, the classical Gaussian kernel has to be left- and right-multiplied by two
extra diagonal matrices.

In this paper we provide a lower-level justification for this specific empirical modification required in
the Gaussian kernel of Divisive Normalization.
Here we assume that the psychophysical behavior described by Divisive Normalization comes from
neural interactions following the Wilson-Cowan equations.
In particular, we identify the Divisive Normalization response with the stationary regime of a Wilson-Cowan model.
From this identification we derive an expression for the Divisive Normalization kernel
in terms of the interaction kernel of the Wilson-Cowan equations. It turns out that the Wilson-Cowan kernel
is left- and-right multiplied by diagonal matrices with high-pass structure.
In conclusion, symmetric Gaussian inhibitory relations between wavelet-like sensors wired in the lower-level Wilson-Cowan model
lead to the appropriate non-symmetric kernel that has to be empirically included in Divisive Normalization
to explain a wider range of phenomena.

\end{abstract}

\section{Introduction}

The general discussion on the circuits and mechanisms underlying the Divisive Normalization addressed in \cite{Carandini12}
suggests that there may be different architectures leading to this specific computation.
Recent results suggest specific mechanisms for Divisive Normalization in certain situations \cite{Carandini16},
but the general debate on the different physiological implementations that may occur is still open.
On the other hand, a number of evidences and functional advantages suggest that the interaction kernel in
Divisive Normalization should be adaptive (i.e. signal or context dependent) \cite{Schwartz09,Schwartz11,Coen12,Coen13}.

Therefore, it is interesting to relate this successful adaptive gain control computation to other
models of interaction in neural populations to explore alternative implementations or new interpretations
of signal dependence in the kernel. An interesting possibility to consider is the classical Wilson-Cowan
model \cite{Wilson72,Wilson73}, which is subtractive in nature.

Subtractive and divisive adaptation models have been qualitatively related \cite{Wilson93,Cowan02}.
Both models have been shown to have similar advantages in information-theoretic terms:
univariate local histogram equalization in Wilson-Cowan \cite{Bertalmio14} and multivariate probability density factorization in Divisive Normalization \cite{Malo10}.
Additionally, both models provide similar descriptions of pattern discrimination \cite{Wilson93,Bertalmio17}.
However, despite all these similarities and relations, no direct analytical correspondence has been established between these models yet.

In this paper, we assume that the psychophysical behavior described by Divisive Normalization comes from
neural interactions following the Wilson-Cowan equations.
In particular, we identify the Divisive Normalization response with the stationary regime of a Wilson-Cowan model.
From this identification we derive an expression for the Divisive Normalization kernel
in terms of the interaction kernel of the Wilson-Cowan equations.
Interestingly, this relation explains (or is consistent with) the sort of empirical modifications that have to
be included ad-hoc in Divisive Normalization based on Gaussian kernels to account for a variety of phenomena.

The structure of the paper is as follows:
in Section \ref{motivation} we recall some results presented in \cite{Martinez17b},
where we showed the ad-hoc modifications required in classical Gaussian kernels in Divisive Normalization
for a proper balance of the different subbands in contrast perception.
In Section \ref{equivalence} we derive the analytical relation between the kernel of Divisive Normalization and the kernel in the Wilson-Cowan equations so that these models are equivalent.
In Section \ref{analysis} we illustrate the elements and the effect of the analytical result using a specific input image.
Finally, in Section \ref{discussion} we discuss the consequences of the result.

\section{Motivation: empirically tuning the Divisive Normalization}
\label{motivation}

\begin{figure}[!b]
	\centering
    \small
    \setlength{\tabcolsep}{2pt}
    \begin{tabular}{c}
    \hspace{-0.0cm} \includegraphics[width=\textwidth]{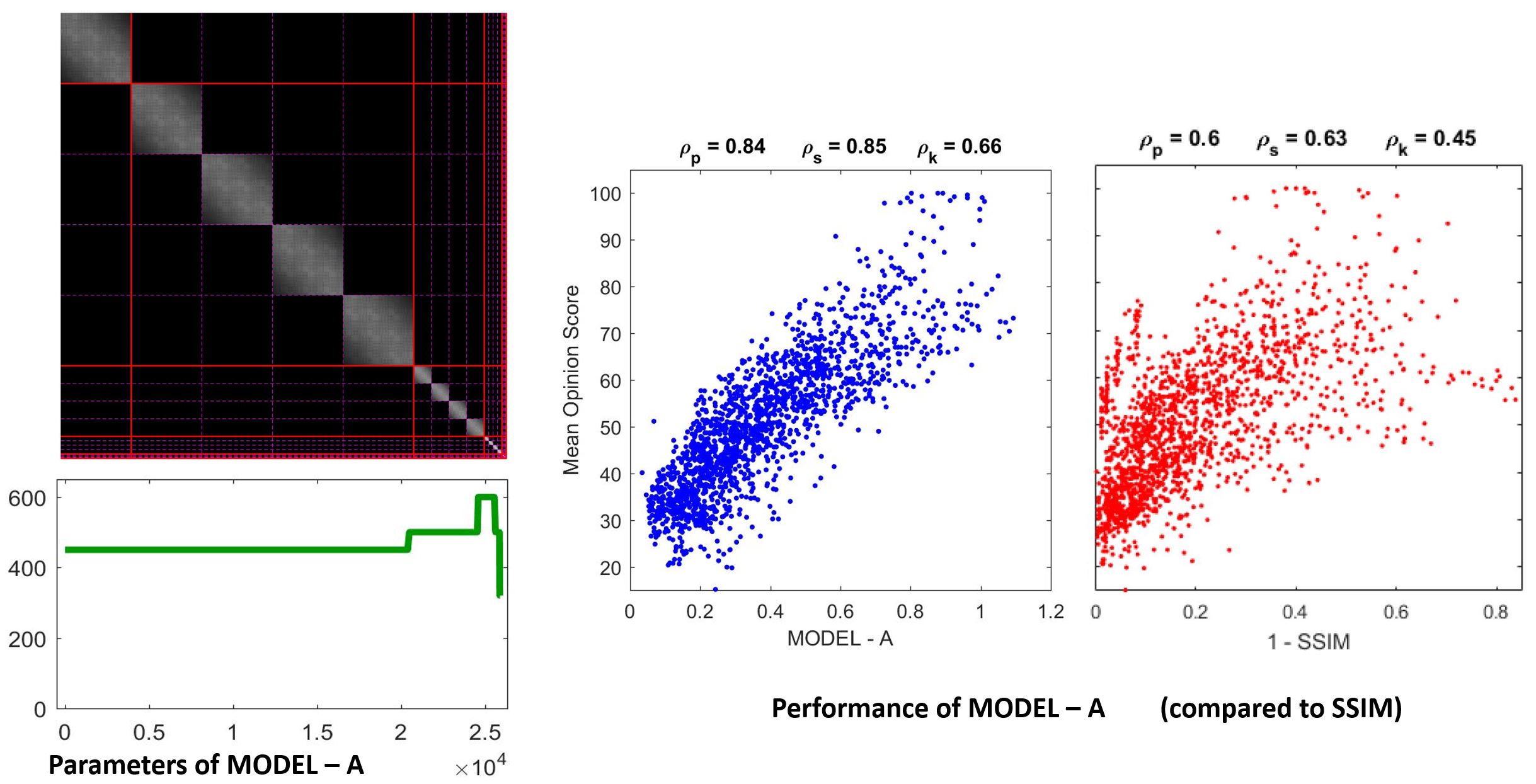} \\
    \end{tabular}
    \vspace{-0.15cm}
	\caption{\emph{Parameters of MODEL-A (left) and performance on large scale naturalistic database (right)}.
    The parameters are: the interaction kernel (matrix on top), and the semisaturation per subband vector (plot as a function of the wavelet coefficient -from high-to-low frequencies-).}
    \label{param_perform_A}
    \vspace{-0.15cm}
\end{figure}

Cascades of Linear+NonLinear Divisive Normalization transforms \cite{Heeger92,Carandini94,Carandini12} can be easily
tuned using the derivatives introduced in \cite{Martinez17a} to reproduce the perception of image distortion in naturalistic environments.
Previous brute-force explorations \cite{Watson02,Malo10,Laparra10a} suggested that spatial interactions in
divisively normalized wavelets are more relevant to reproduce subjective opinion than scale
and orientation interactions.
Good results obtained from optimization of such spatial-only kernels confirms this \cite{Martinez17a}.
In this intraband-only Divisive Normalization the vector of V1-like activations, $\vect{x}$, depends on the energy of linear wavelet
responses, $\vect{e}$, dimension-wise normalized by a sum of neighbor energies,
\vspace{-0.0cm}
\begin{equation}
      \vect{x} = \frac{\vect{e}}{\vect{b} + H^{\vect{p}} \cdot \vect{e}} = \mathbb{D}^{-1}_{\left( \vect{b} + H^{\vect{p}} \cdot \vect{e} \right)} \cdot \vect{e}
      \label{DNormA}
\end{equation}
where the kernel $H^{\vect{p}}$ only considers the departure in spatial position, $\Delta \vect{p}$, between sensors of the same subband.
The division in Eq.~\ref{DNormA} is a Hadamard (element-wise) operation. Further computations are more intuitive if these Hadamard operations are substituted by regular products of diagonal matrices on vectors: $\vect{a} \odot \vect{b} = \mathbb{D}_{\vect{a}} \cdot {\vect{b}}$, where $\mathbb{D}_{\vect{a}}$ is a diagonal matrix with vector $\vect{a}$ in the diagonal \cite{Minka00}. This is what leads to the (more convenient) matrix form of Divisive Normalization in the right-hand side of Eq. \ref{DNormA} \cite{Martinez17a}.
We will refer to this intraband-only model as \textbf{Model A}.
Fig~\ref{param_perform_A} shows the intraband kernel and semisaturation for the Divisive Normalization to reproduce perceived distortion in naturalistic databases.
\vspace{-0.0cm}

\paragraph{Obvious limitations of intraband kernels.}
Despite successful optimization of \textbf{Model A} over large naturalistic image quality databases \cite{Martinez17a},
some basic effects with artificial stimuli may be poorly reproduced \cite{Martinez17b}:
while \textbf{Model A} explains cross-orientation and cross-scale masking for low frequency tests seen on high frequency backgrounds
it is not the case the other way around.
Fig~\ref{failure_A} shows series of synthetic data that illustrate the nonlinear and adaptive nature of contrast responses.
Moreover, it shows the failures of \textbf{Model A} for high frequency tests.
To fix this, a more balanced interaction between subbands in the denominator of Eq. \ref{DNormA} is required, \emph{which cannot be introduced in intraband-only kernels}.
\vspace{-0.0cm}

\begin{figure}[!t]
	\centering
    \small
    \setlength{\tabcolsep}{2pt}
    \begin{tabular}{c}
    \hspace{0.5cm} \includegraphics[width=0.7\textwidth]{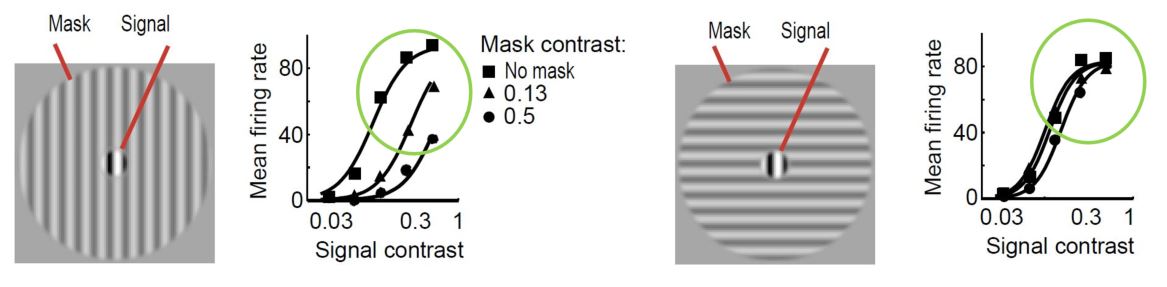} \\[0.1cm]
    \hline \\[0.1cm]
    \hspace{-0.0cm} \includegraphics[width=0.75\textwidth]{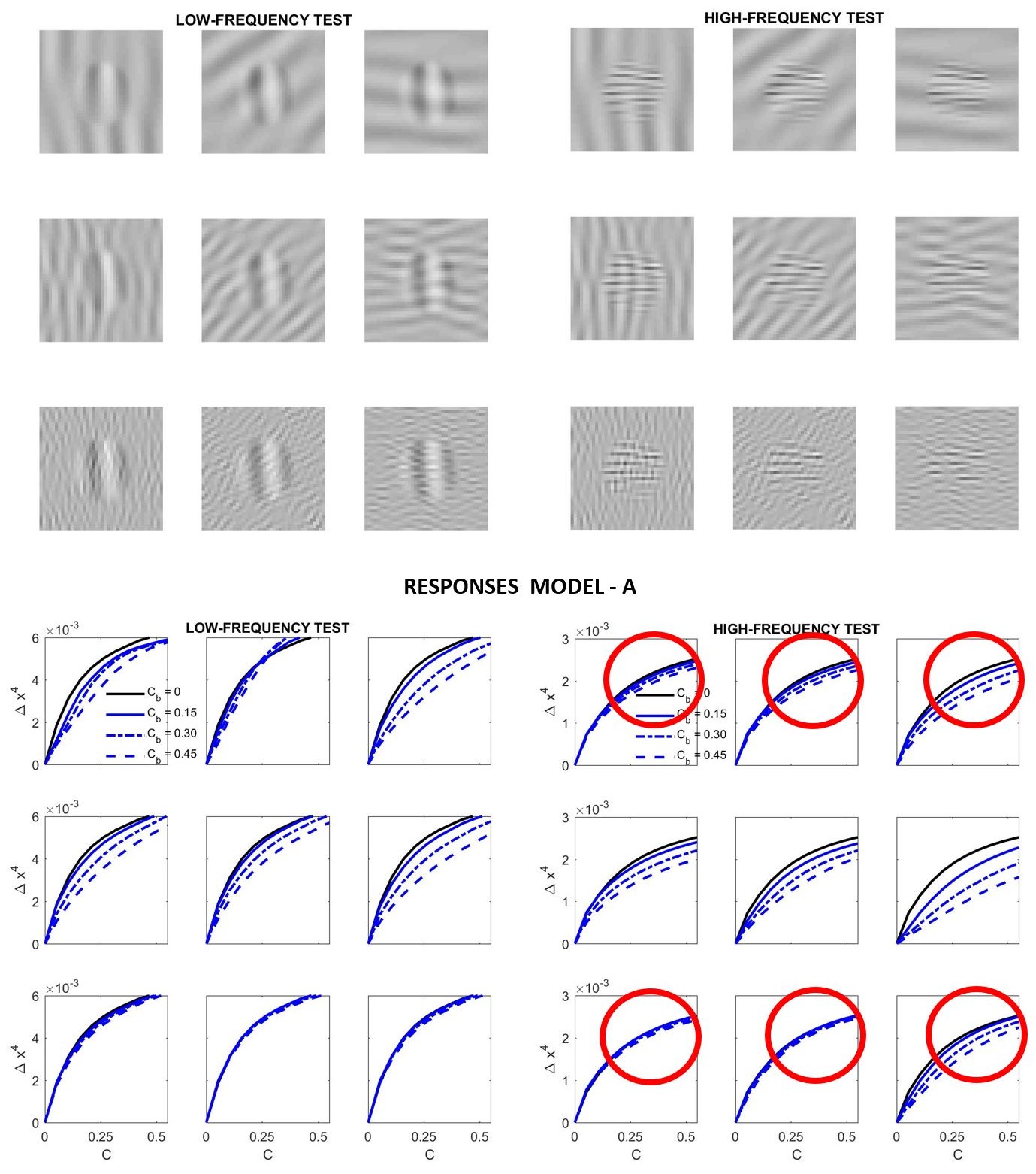} \\
    \end{tabular}
    \vspace{-0.15cm}
	\caption{Experimental response of V1 neurons (mean firing rate) in masking situations. Adapted from Schwartz and Simoncelli (2001); Cavanaugh (2000).
It is important to stress the decay in the response when test and mask do have the same spatio-frequency characteristics, as opposed to the case where they do
not (difference in the circles in green).
\emph{Relative success and failures of optimized model}.
    Model-related construction of stimuli simplify the reproduction of results form model outputs and straightforward interpretation of results.
}\label{failureA}
    \vspace{-0.15cm}
\end{figure}

\begin{figure}[!t]
	\centering
    \small
    \setlength{\tabcolsep}{2pt}
    \begin{tabular}{c}
    \hspace{-1cm} \includegraphics[width=1.1\textwidth]{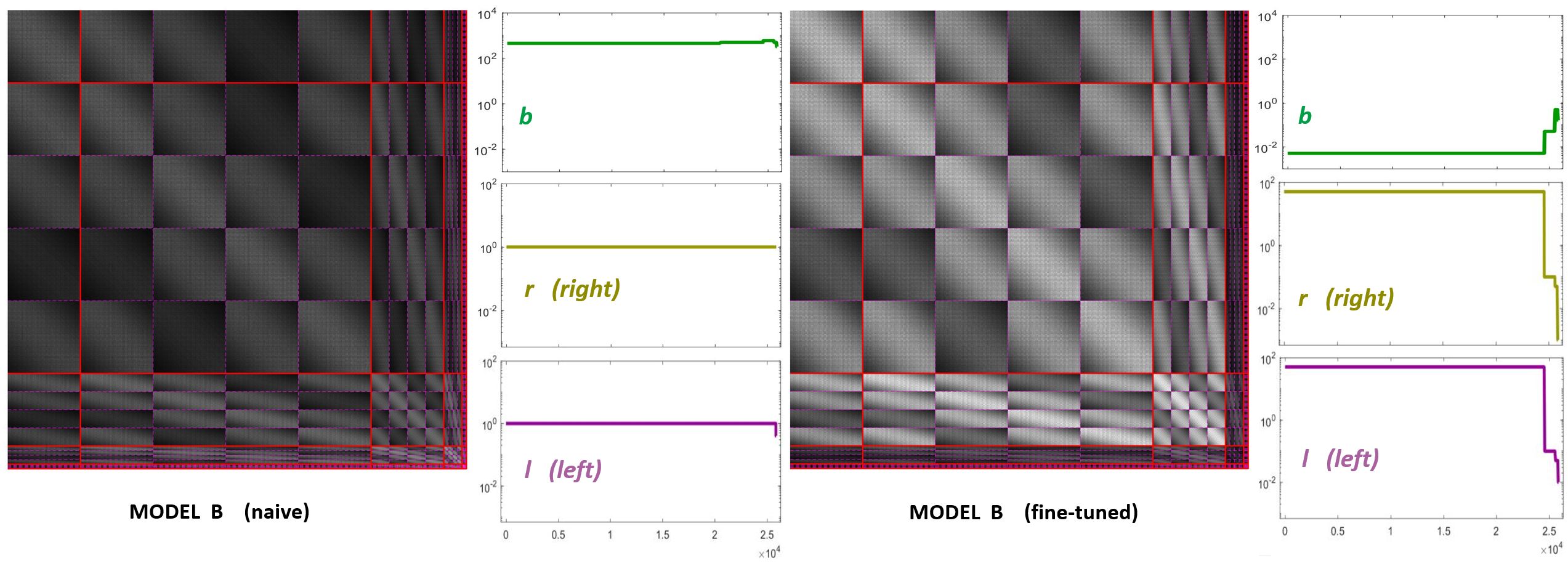} \\
    \end{tabular}
    \vspace{-0.15cm}
	\caption{\emph{Parameters of the modified models}.
    Left panel shows the interaction matrix and the semisaturation vector of the first guess for Model - B. It is called \emph{naive} because the semisaturation and amplitudes of the kernel are imported from the optimized case. The panel at the right shows the corresponding parameters for the fine-tuned version of Model B.
}\label{new_parameters}
    \vspace{-0.15cm}
\end{figure}

\begin{figure}[!t]
	\centering
    \small
    \setlength{\tabcolsep}{2pt}
    \begin{tabular}{c}
    \hspace{-1.0cm} \includegraphics[width=1.1\textwidth]{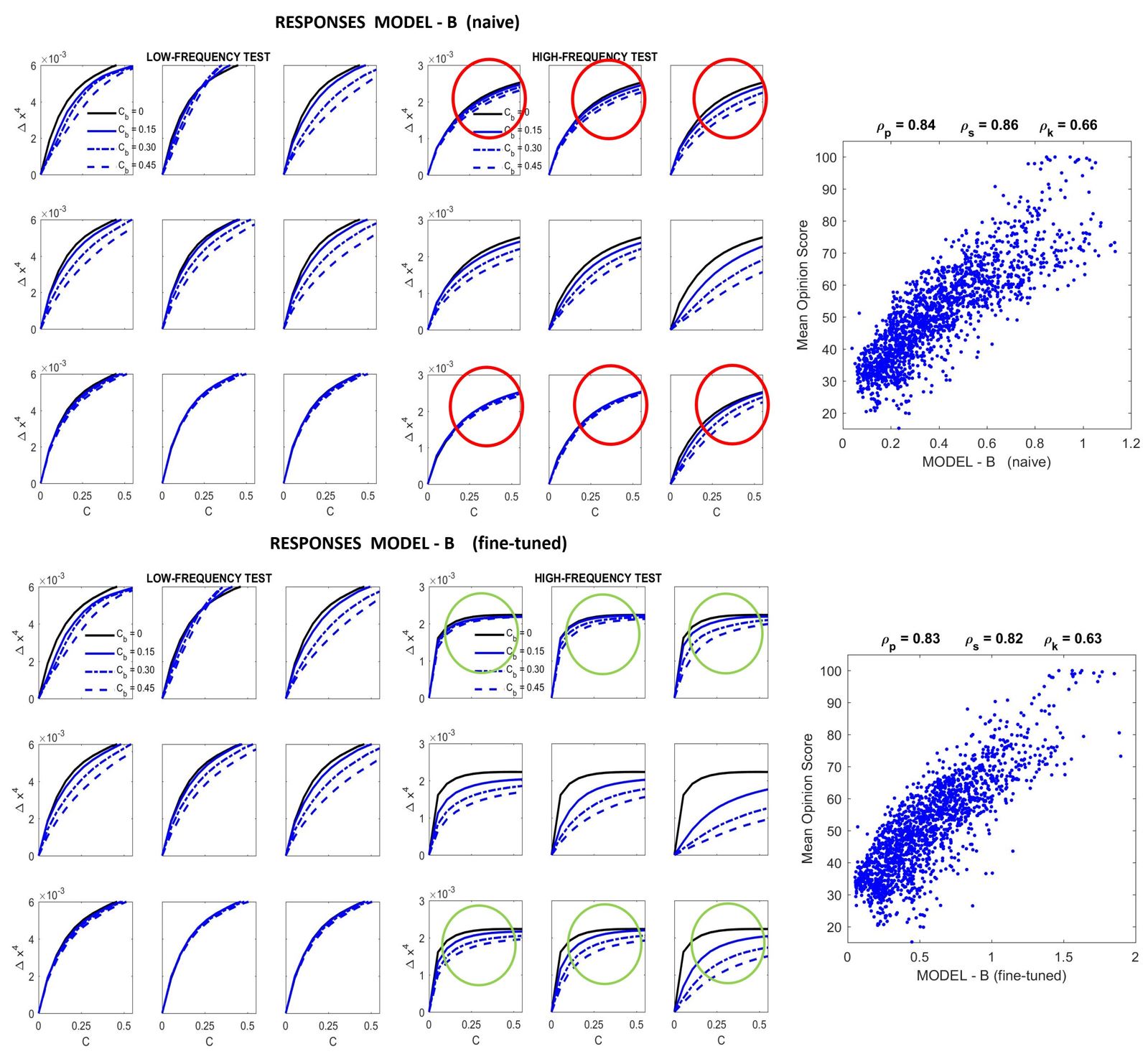} \\
    \end{tabular}
    \vspace{-0.15cm}
	\caption{\emph{Responses for artificial stimuli (left) and performance in the natural image database (right) of the naive model (top) and fine-tuned model (down)}.
}\label{successB}
    \vspace{-0.15cm}
\end{figure}

\paragraph{Solution goes beyond Watson \& Solomon kernels.}
The first guess to fix the imbalance is substituting the spatial-only kernel $H^{\vect{p}}$ in Eq. \ref{DNormA}
by more general kernels, as the one proposed by Watson \& Solomon, $H^{ws} = H^{\vect{p}} \odot H^{f} \odot H^{\phi}$,
that not only depends on departures in position, $\vect{p}$, but also in frequency, $f$, and in orientation $\phi$ \cite{Watson97}.
We will call this first guess for correction \textbf{Model B - naive}.
However, it turns out that
Gaussian $H^{ws}$ may not provide the appropriate balance either:
low frequency backgrounds may still have too much energy and bias the result for high frequency tests.
In \cite{Martinez17b} we showed that this may be fixed ad-hoc by \emph{left} and \emph{right} multiplication of the Watson \& Solomon kernel with extra diagonal matrices:
\begin{equation}
      H = \mathbb{D}_{\vect{l}} \cdot H^{ws} \cdot \mathbb{D}_{\vect{r}}
      \label{new_kernel_eq}
\end{equation}
While $\mathbb{D}_{\vect{r}}$, pre-weights the subbands of $\vect{e}$ to moderate the effect of low frequencies before computing the interaction,
$\mathbb{D}_{\vect{l}}$, tunes the relative weight of the masking for each sensor, moderating low frequencies again.
Additionally to the changes in $H$ to account for the artificial stimuli,
the models B included an extra constant to keep the output dynamic range as in the simpler model of Eq.~\ref{DNormA},
just to keep the good performance of \textbf{Model A} for naturalistic stimuli.
We will refer to this empirically-tuned model to as \textbf{Model B - fine-tuned}.
Fig~\ref{new_parameters} compares the parameters of Model B - naive and Model B - fine-tuned, Fig~\ref{successB} shows how the fine-tuning solves
the problem for artificial stimuli and how this fine-tuning preserves the good behavior on the naturalistic database.

\paragraph{Question: where the fine-tuned solution comes from?.}
Summarizing, in order to account simultaneously for the perception of distortion in naturalistic databases
and for contrast perception of synthetic stimuli the response of \textbf{Model B - fine-tuned} is:
\begin{equation}
    \vect{x} = \mathbb{D}_{\vect{k}} \cdot \mathbb{D}^{-1}_{\left( \vect{b} + H \cdot \vect{e} \right)} \cdot \vect{e}
    \label{DN_B}
\end{equation}
where the interaction kernel $H$ requires a specific structure, i.e Eq.~\ref{new_kernel_eq},
and vectors $\vect{l}$ and $\vect{r}$ have high-pass nature (see Fig. 3).
\vspace{0.0cm}

The question is: \emph{where the structure in Eq. \ref{new_kernel_eq} comes from?}.
\vspace{-0.0cm}

In order to explain this structure here we hypothesize that the psychophysical behavior described by Divisive Normalization
comes from (lower-level) neural interactions in V1 (e.g. the classical Wilson-Cowan equations \cite{Wilson72,Wilson73}).
In particular, we identify the Divisive Normalization response with the stationary regime of a Wilson-Cowan model.
From this identification in Section \ref{equivalence} we derive a novel expression for the Divisive Normalization kernel
in terms of the interaction kernel of the Wilson-Cowan equations. As shown below, this will
explain the structure in Eq. \ref{new_kernel_eq}.

\section{Equivalence between Divisive Normalization and Wilson-Cowan}
\label{equivalence}

The Divisive Normalization model \cite{Carandini94,Carandini12} and the Wilson-Cowan model \cite{Wilson72,Wilson73} are alternative (divisive versus subtractive) formulations of the interactions among sensors in neural populations.

In this work we assume that the psychophysical behavior described by the Divisive Normalization response is the stationary solution of the dynamic system defined by the Wilson-Cowan equations,
which leads to a relation between the parameters that describe the interaction and the auto-saturation in these models.

This relation is relevant because it explains the kind of empirical modifications that had to be introduced ad-hoc in \cite{Martinez17b} in the standard Gaussian kernels of Divisive Normalization to simultaneously reproduce the perception of distortions on naturalistic and artificial environments.

\subsection{Modelling cortical interactions.}
In the case of the V1 cortex, we refer to the set of responses of a population of simple cells as the vector $\vect{y}$.
The considered models (Divisive Normalization and Wilson-Cowan) define a nonlinear mapping that transforms the input vector $\vect{y}$ (before the interaction among neurons) into the output vector $\vect{x}$ (after the interaction),
\vspace{-0.2cm}
\begin{equation}
  \xymatrixcolsep{2pc}
  \xymatrix{ \vect{y}  \,\,\,\, \ar@/^0.7pc/[r]^{\scalebox{0.85}{$\mathcal{N}$}} & \,\,\,\, \vect{x}
  }
  \label{global_response}
\end{equation}
In this setting, responses are called \emph{excitatory} or \emph{inhibitory}, depending on the corresponding \emph{sign} of the signal: $\vect{y} = \textrm{sign}(\vect{y}) |\vect{y}| $, and $\vect{x} = \textrm{sign}(\vect{x}) |\vect{x}|$.
The map $\mathcal{N}$ is an adaptive saturating transform, but it preserves the sign of the responses (i.e. $\textrm{sign}(\vect{x})=\textrm{sign}(\vect{y})$).
Therefore, the models care about cell activation (the modulus $|\cdot|$) but not about the excitatory or inhibitory nature of the sensors (the $\textrm{sign}(\cdot)=\pm$).

We will refer to as the \emph{energy} of the input responses to the vector $\vect{e} = |\vect{y}|^\gamma$, where this is an element-wise exponentiation of the amplitudes $|y_i|$.

Given the sign-preserving nature of the nonlinear mapping, for the sake of simplicity in notation, in the rest of the paper the variables $\vect{y}$ and $\vect{x}$ refer to the activations $|\vect{y}|$ and $|\vect{x}|$.

\vspace{-0.0cm}
\subsection{The Divisive Normalization model}
\paragraph{Forward transform.} The input-output transform in the Divisive Normalization is given by Eq.~\ref{DN_B}:
the output vector of nonlinear activations in V1, $\vect{x}$, depends on the energy of the input linear wavelet responses, $\vect{e}$,
which are dimension-wise normalized by a sum of neighbor energies.
The normalization by $\vect{b} + H \cdot \vect{e}$ and the non-diagonal nature of the interaction kernel $H$
implies that some specific response will be attenuated if the activity of the neighbor sensors is high.
Each row of the kernel $H$ describes how the energies of the neighbor sensors attenuate the activity of each sensor after the interaction.
The each element of the vectors $\vect{b}$ and $\vect{k}$ respectively determine the semisaturation and the dynamic range of the nonlinear response
of each sensor.

\vspace{-0.0cm}
\paragraph{Inverse transform.} Relation between the two models is easier to obtain by identifying the corresponding decoding transforms in both models.
In the case of Divisive Normalization, the analytical inverse is \cite{Malo06a,Martinez17a}:
\begin{equation}
      \vect{e} = \left( I - \mathbb{D}^{-1}_{\vect{k}}\cdot\mathbb{D}_{\vect{x}}\cdot H \right)^{-1} \cdot \mathbb{D}_{\vect{b}} \cdot \mathbb{D}^{-1}_{\vect{k}} \cdot \vect{x}
      \label{invDN}
\end{equation}

\vspace{-0.0cm}
\subsection{The Wilson-Cowan model}
\paragraph{Dynamical system.} In the Wilson-Cowan model the variation of the activation vector, $\vect{\dot{x}}$, increases with the energy of the input, $\vect{e}$, but, for each sensor, this variation is also moderated by its own activity and by a linear combination
of the activities of the neighbor sensors,
\vspace{-0.0cm}
\begin{equation}
      \vect{\dot{x}} = \vect{e} - \mathbb{D}_{\vect{\alpha}} \cdot \vect{x} - \vect{W} \cdot f(\vect{x})
      \label{EqWC}
\end{equation}
where $\vect{W}$ is the matrix that describes the damping factor between sensors, and $f(\vect{x})$ is a dimension-wise saturating nonlinearity. Different convenient approximations can be taken for that saturation (either piece-wise or continuous, see fig. \ref{f_x}) as for instance (a) $f(\vect{x}) \approx \vect{x}$, or (b) $f(\vect{x}) \approx \vect{x}^\beta$.

Note that in Eq.~\ref{EqWC} both the inhibitory and the excitatory cells are included in the same vector, thus the two traditional Wilson-Cowan equations are represented here by a single expression.


\paragraph{Steady state and inverse.} Under the approximation (a) of the saturation, the stationary solution of the above differential equation, $\vect{\dot{x}} =0$ in Eq.~\ref{EqWC}, leads to the following decoding (input-from-output) relation:
\begin{equation}
      \vect{e} = \left( \mathbb{D}_{\vect{\alpha}} + \vect{W} \right) \cdot \vect{x}
      \label{invWC}
\end{equation}

\begin{Figure}
 \centering
 \small
 \includegraphics[width=4cm,height=3.5cm]{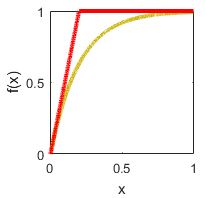}
 \captionof{figure}{\emph{\textbf{Saturating function in Wilson-Cowan.}
Sensible choices for this function include piece-wise linear functions (example in red) or saturating exponentials (example in orange).}}
             \label{f_x}
\end{Figure}

\vspace{-0.0cm}
\subsection{Analytical relation between models}

In this work we derive the equivalence between the models through the identification between the decoding equations in both
cases (Eqs. \ref{invDN} and \ref{invWC}). This allows us to establish a relation between the
parameters that describe the interaction and the auto-attenuation in
both models.
The identification is simpler by taking the series expansion of the inverse in Eq. \ref{invDN}.
This expansion was used in \cite{Malo06a} because it clarifies the condition for invertibility of Divisive Normalization:
\begin{equation}
      \left( I - \mathbb{D}^{-1}_{\vect{k}} \cdot \mathbb{D}_{\vect{x}} \cdot H \right)^{-1}
      = I + \sum_{n=1}^{\infty} \left( \mathbb{D}^{-1}_{\vect{k}} \cdot \mathbb{D}_{\vect{x}} \cdot H \right)^n
      \nonumber
\end{equation}
The inverse exist if the eigenvalues of $\mathbb{D}^{-1}_{\vect{k}} \cdot \mathbb{D}_{\vect{x}} \cdot H$ are smaller than one so that the series converges.
In fact if the eigenvalues are small, the inverse can be well approximated by a small number of terms in the series.
\begin{eqnarray}
      \vect{e} & = & \mathbb{D}_{\vect{b}} \cdot \mathbb{D}^{-1}_{\vect{k}} \cdot \vect{x} + \left( \mathbb{D}^{-1}_{\vect{k}} \cdot \mathbb{D}_{\vect{x}} \cdot \vect{H} \right) \cdot \mathbb{D}^{-1}_{\vect{k}} \cdot \mathbb{D}_{\vect{b}} \cdot \vect{x} + \nonumber \\
      & & + \left( \mathbb{D}^{-1}_{\vect{k}} \cdot \mathbb{D}_{\vect{x}} \cdot \vect{H} \right)^2 \cdot \mathbb{D}^{-1}_{\vect{k}} \cdot \mathbb{D}_{\vect{b}} \cdot \vect{x} + \nonumber \\
      & & + \left( \mathbb{D}^{-1}_{\vect{k}} \cdot \mathbb{D}_{\vect{x}} \cdot \vect{H} \right)^3 \cdot \mathbb{D}^{-1}_{\vect{k}} \cdot \mathbb{D}_{\vect{b}} \cdot \vect{x} + \cdots \nonumber \\[0.4cm]
      \vect{e} &\approx& \left( \mathbb{D}_{\vect{b}} \cdot \mathbb{D}^{-1}_{\vect{k}} + \mathbb{D}^{-1}_{\vect{k}} \cdot \mathbb{D}_{\vect{x}} \cdot \vect{H} \cdot \mathbb{D}_{\vect{b}} \cdot \mathbb{D}^{-1}_{\vect{k}} \right) \cdot \vect{x}
      \label{approx_invDN}
\end{eqnarray}
Now, identification of Eq. \ref{approx_invDN} and \ref{invWC} (i.e. the first order approximation of the inverse of the Divisive Normalization, and the decoding assuming a piece-wise linear approximation of $f(\vect{x})$ in the Wilson-Cowan model) is straightforward. As a result, we get the following relations between the parameters of both models:
\begin{eqnarray}
      \vect{b} &=& \vect{k} \odot \vect{\alpha} \nonumber \\
      H &=& \mathbb{D}_{\left(\frac{\vect{k}}{\vect{x}}\right)} \cdot \vect{W} \cdot \mathbb{D}_{\left(\frac{\vect{k}}{\vect{b}}\right)}
      \label{relation_W_H}
\end{eqnarray}
Note that the resulting $H$ in Eq. \ref{relation_W_H} has exactly the structure that had to be introduced ad-hoc in Eq.~\ref{new_kernel_eq}.
Both models are equivalent if the Divisive Normalization kernel inherits the structure from the Wilson-Cowan kernel modified by the these pre- and post- diagonal matrices.
Note that the weights after the interaction (the diagonal matrix at the left) is signal dependent, which implies that the interaction kernel in Divisive Normalization should be adaptive.
In the next Section we show that the vectors (Hadamard quotients) $\vect{k}/\vect{x}$ and $\vect{k}/\vect{b}$ do have the high-pass frequency nature
that explains why the low frequencies in $\vect{e}$ had to be attenuated by $\vect{r}$ and $\vect{l}$.

\section{Analysis of the equivalence}
\label{analysis}

In this section we consider an illustrative signal and sensible values for the parameters involved in Eq. \ref{relation_W_H} ($\vect{k}$, $\vect{b}$ and $\vect{W}$),
to analyze the effects in the Divisive Normalization kernel and compare with the hand-crafted kernel in Eq. \ref{new_kernel_eq}.

Here we compare the empirical filters (vectors $\vect{l}$ and $\vect{r}$) presented in Section 1 (Eq. \ref{new_kernel_eq}) with the corresponding vectors in Eq. \ref{relation_W_H}.
We also compare the masking term in the denominator of Divisive Normalization using (1) the Gaussian kernel $H^{ws} \cdot \vect{e}$, (2) the empirically modified kernel $\mathbb{D}_{\vect{l}} \cdot H^{ws} \cdot \mathbb{D}_{\vect{r}} \cdot \vect{e}$, and (3) the theoretically derived kernel obtained from Eq. \ref{relation_W_H}, $\mathbb{D}_{\left(\frac{\vect{k}}{\vect{x}}\right)} \cdot \vect{W} \cdot \mathbb{D}_{\left(\frac{\vect{k}}{\vect{b}}\right)} \cdot \vect{e}$.
In this comparison we assume a Gaussian wiring in $\vect{W}$.

\begin{figure}[!t]
	\centering
    \small
    \setlength{\tabcolsep}{2pt}
    \begin{tabular}{c}
    \hspace{-2.0cm} \includegraphics[width=1.2\textwidth]{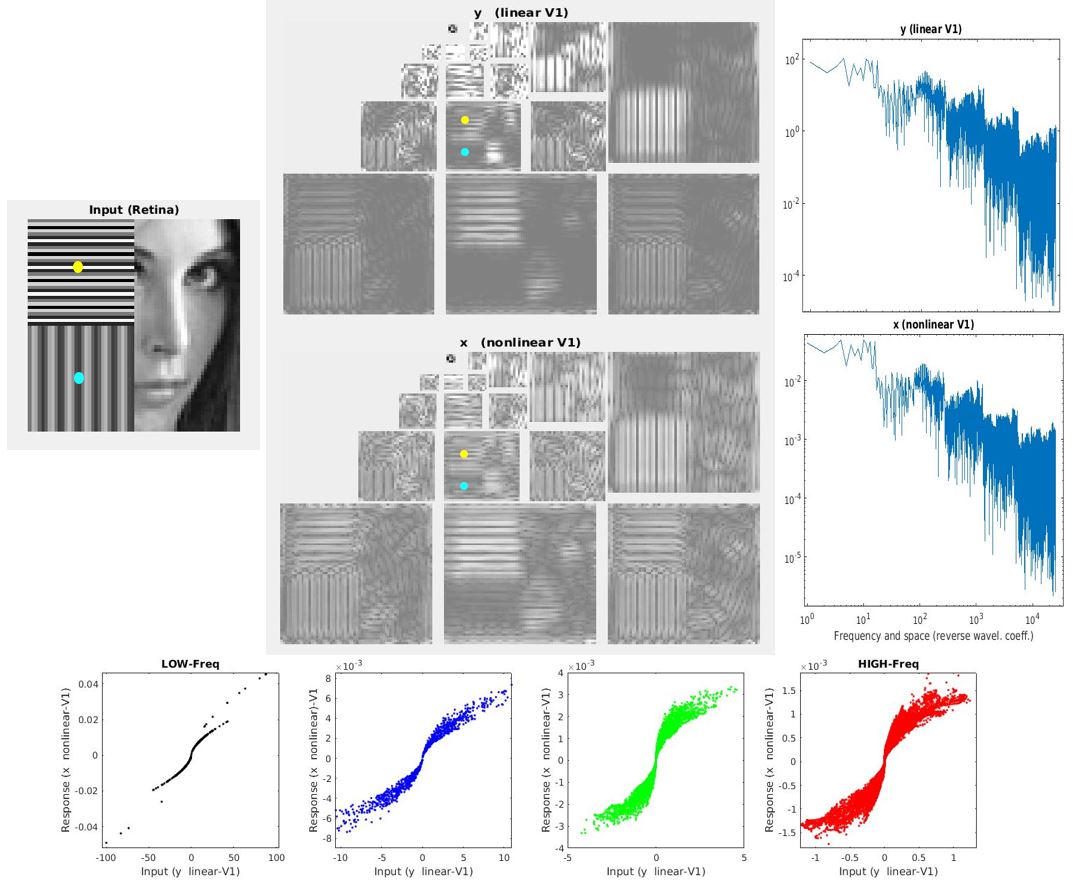} \\
    \end{tabular}
    \vspace{-0.15cm}
	\caption{\emph{Responses of the model in V1: localized oriented filters at different scales}.
     Natural images typically have low-pass responses.
     Divisive Normalization implies adaptive saturating nonlinearity depending on the neighbors (i.e. a family of sigmoid functions in the input-output scatter plots).
     Note the response of the medium frequency horizontal sensors tuned to different positions highlighted in yellow and blue.
}\label{explanation}
    \vspace{-0.15cm}
\end{figure}

\begin{figure}[!t]
	\centering
    \small
    \setlength{\tabcolsep}{2pt}
    \begin{tabular}{c}
    \hspace{-0.0cm} \includegraphics[width=0.8\textwidth]{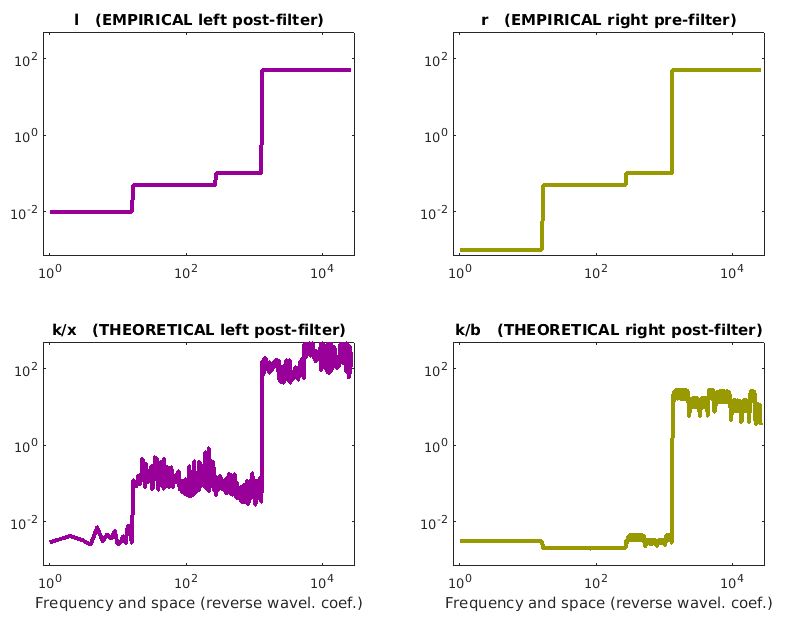} \\
    \end{tabular}
    \vspace{-0.15cm}
	\caption{\emph{Vectors in the left- and right- diagonal matrices that multiply the Gaussian kernel in the empirical tuning represented by Eq. \ref{new_kernel_eq} (top) and in the theoretically derived Eq. \ref{relation_W_H} (bottom)}.
}\label{filters}
    \vspace{-0.15cm}
\end{figure}

\begin{figure}[!t]
	\centering
    \small
    \setlength{\tabcolsep}{2pt}
    \begin{tabular}{c}
    \hspace{-0.0cm} \includegraphics[width=0.7\textwidth]{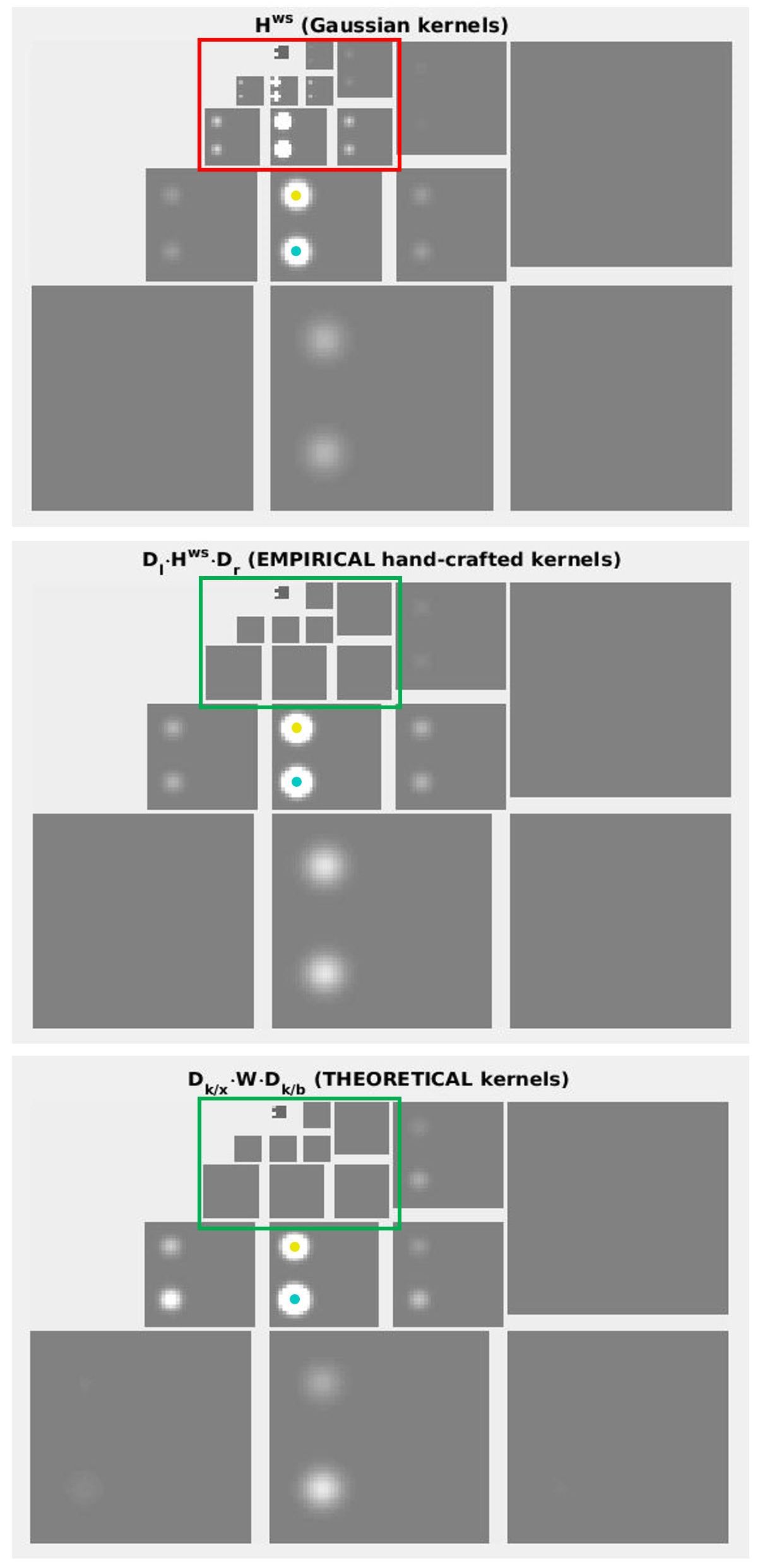} \\
    \end{tabular}
    \vspace{-0.15cm}
	\caption{\emph{Interaction kernels for the sensors highlighted in yellow and blue in Fig. \ref{explanation}}. In the Gaussian case (top) both low and high frequencies symmetrically affect each sensor. This implies the higher energy of low-frequency components bias the response and ruins the masking curves. This was corrected ad-hoc using right- and left- multiplication in Eq. \ref{new_kernel_eq} by hand-crafted high-pass filters. This leads to the empirical kernels in the center. In both cases (Gaussian and hand-crafted, top and center) the size of the interaction neighborhood is signal independent (the same for both locations).
The kernels at the bottom are those obtained from Eq. \ref{relation_W_H}. These theoretically-derived kernels remove the bias due to the low frequencies (just as the hand-crafted kernel), but also introduce an extra signal dependence: note that the interaction neighborhood now depends on the location (bigger for the sensor in blue) due to the different value of the signal.
}\label{kernels}
    \vspace{-0.15cm}
\end{figure}

\begin{figure}[!t]
	\centering
    \small
    \setlength{\tabcolsep}{2pt}
    \begin{tabular}{c}
    \hspace{-0.0cm} \includegraphics[width=0.7\textwidth]{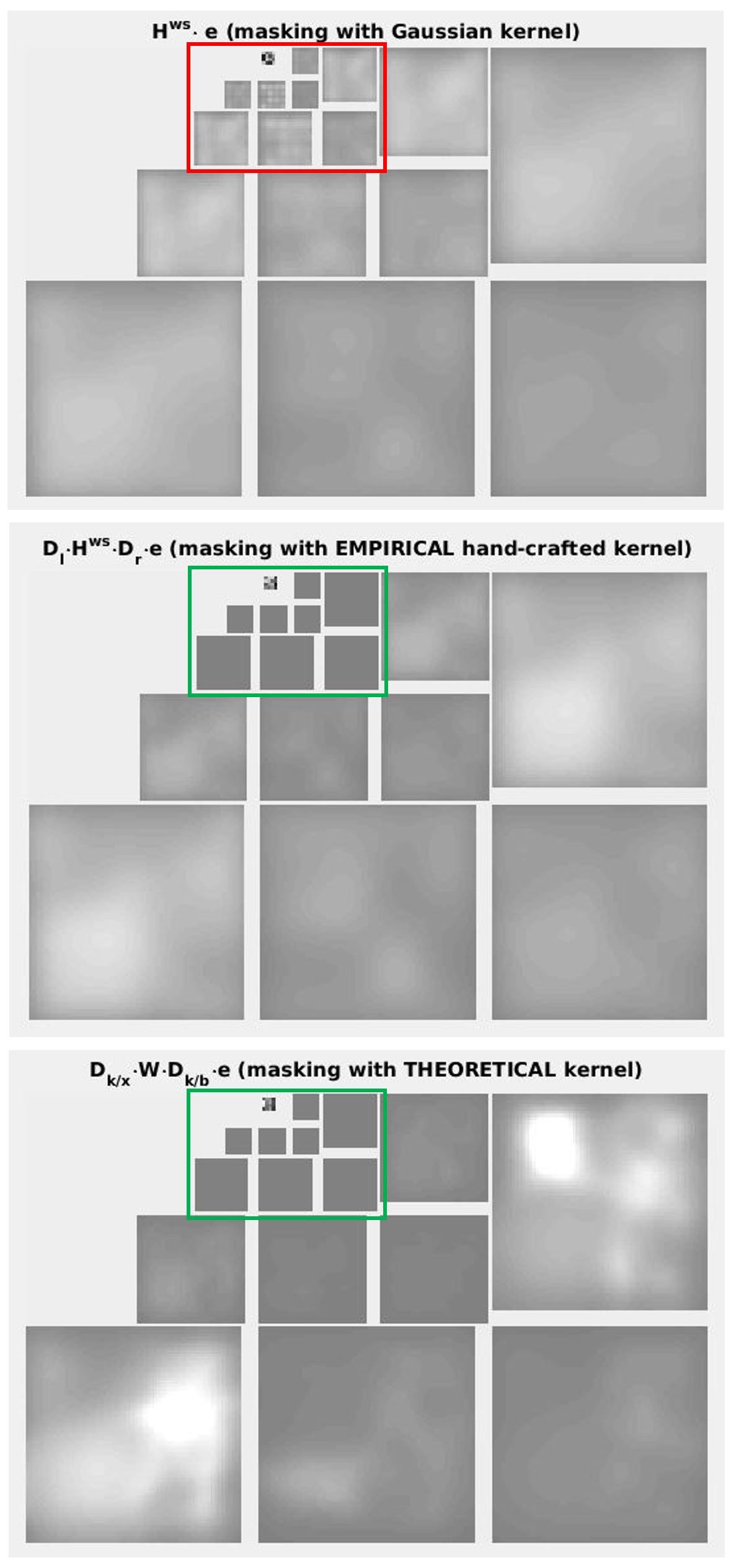} \\
    \end{tabular}
    \vspace{-0.15cm}
	\caption{\emph{Masking term in the denominator of the Divisive Normalization for various kernel choices.} Gaussian kernel (top), Empirically tuned hand-crafted kernel (center), and Theoretically derived kernel (bottom)}\label{masking}
    \vspace{-0.15cm}
\end{figure}

Before doing so, given an illustrative input image,
Fig.\ref{explanation} shows the corresponding responses of linear and nonlinear V1-like sensors based on steerable wavelets.
Typical responses for natural images are low-pass signals.
Fig.\ref{filters} compares the empirical vectors with those based on the relation with the Wilson-Cowan model: both show similar high-pass nature.
Fig. \ref{kernels} compares different individual interaction kernels (rows of the kernel matrix) for the two sensors highlighted in yellow and blue in Fig. \ref{explanation}.
Fig. \ref{masking} compares the masking terms: the one derived from the proposed relation is more similar to the (more general) empirically tuned result than
to the one based on the (more limited) Gaussian kernel.

\section{Final remarks}
\label{discussion}

This relation between models has a range of consequences.
First, assuming fixed (hard-wired) interaction between the sensors in the Wilson-Cowan model,
Eq.~\ref{relation_W_H} implies that the required kernel in Divisive Normalization, $H$, not only inherits the wiring in $\vect{W}$, but it also should be signal-dependent.
Second, functional forms depending on \emph{proximity} (as in the Watson-Solomon kernel $H^{ws}$)
seem sensible choices for wiring in $\vect{W}$, which would justify the hand-crafted trick in Eq. \ref{new_kernel_eq}.
Last, but more importantly, Eq.~\ref{relation_W_H} implies that the variety of dynamic analysis already done for Wilson-Cowan systems \cite{Sejnowski09} can also be applied to the wide range of phenomena described by Divisive Normalization.


\section*{References}

\bibliographystyle{unsrt}

\renewcommand{\baselinestretch}{1}
{\footnotesize
\bibliography{the_WC_DN}}
\end{document}